\documentclass[aps,prb,twocolumn]{revtex4}

\usepackage{amsmath}
\usepackage{color}
\usepackage{amssymb}

\usepackage{graphicx}

\begin{document}

\title{Temperature-dependent Ginzburg-Landau parameter}
\author{Vladimir P. Mineev and Vincent P. Michal}
\affiliation{Service de Physique Statistique, Magn\'{e}tisme et
Supraconductivit\'{e}, Institut Nanosciences et Cryog\'{e}nie, UMR-E CEA/UJF-Grenoble1, F-38054 Grenoble, France}

\date{\today}
\begin{abstract}
Taking into account both the orbital and the paramagnetic depairing  effects we derive  a  simple analytic formula for the temperature dependence of the Ginzburg-Landau parameter $\kappa$
valid in vicinity of field dependent critical temperature
in a type-II superconductor.

\smallskip


\end{abstract}

\maketitle

\section {Introduction} It is usually considered that the Ginzburg-Landau parameter (the ratio $\kappa=\lambda/\xi$ of the London penetration depth
$\lambda$ and the coherence length $\xi$) is temperature independent.
The effective tool for the experimental determination of the Ginzburg-Landau parameter near the critical temperature is  given by the famous Abrikosov formula\cite{Abrikosov} for the  field derivative of magnetization near the upper critical field in the type-II superconductors
\begin{equation}
\left.\frac{dM}{dH}\right|_{H=H_{c2}}=\frac{1}{4\pi(2\kappa^2-1)\beta_A}.
\label{Abr}
\end{equation}
Here $\beta_A=\langle|\Delta|^4\rangle/(\langle|\Delta|^2\rangle)^2$ is the Abrikosov parameter.
In practice, it is convenient  to  use the Ehrenfest formula which relates the slope of magnetization curve near $H_{c2}$ to the specific heat jump at $T=T_c(H_{c2})$
\begin{equation}
\frac{\Delta C}{T}=\left ( \frac{dH_{c2}}{dT}\right )^2\left (\frac{dM}{dH}  \right ).
\label{Ehrenfest}
\end{equation}
From Eqs. (\ref{Abr}) and (\ref{Ehrenfest}) we obtain 
\begin{equation}
\kappa=\frac{1}{\sqrt{2}}\sqrt{1+\frac{T}{4\pi\beta_A\Delta C}\left (\frac{dH_{c2}}{dT}\right )^2}
\label{GLfull}
\end{equation}
For strong type two superconductors with $\kappa>>1$
\begin{equation}
\kappa\approx\left |\frac{d H_{c2}}{d T}\right |\sqrt{\frac{T}{8\pi\beta_A\Delta C}}.
\label{GL}
\end{equation}
The slope of the upper critical field in the Ginzburg-Landau region is temperature-independent. The same is true for the ratio $T/\Delta C$. However, experimentally in several heavy fermionic compounds there was revealed a fast drop of the GL parameter with decreasing temperature.\cite{Ikeda, Haga} So, the GL parameter proved to be a function of temperature. 
Already in the earliest experimental study there was suggested\cite{Ikeda} that this temperature dependence is introduced by the Zeeman depairing effect. Then the temperature dependence of GL parameter has been  discussed theoretically in the paper \cite{Adachi} making use  the numerical solution of the Eilenberger equations taking into account both the orbital and the paramagnetic effect. Later the analytical expression for the GL parameter  below $T_c(H)$  in the limit of strong paramagnetic effect has been found in the paper\cite{Houzet}. In spite of these results it is still of definite interest to give a simple analytic formula for Gl parameter valid near the phase transition line  taking into account both the paramagnetic and the orbital depairing.
Here we find this dependence in a straightforward way. Then we compare this with result of paper \cite{Houzet}.

We note the temperature-dependent Ginzburg-Landau parameter $\kappa(T)$ instead of $\kappa_2$ used in the experimental literature and reserve the term Maki parameter $\alpha_M$ for the ratio of orbital critical field to the paramagnetic limiting field at zero temperature $\sqrt{2}H_{c20}/H_p$.  In  a clean superconductor $\alpha_M\approx T_c/mv_F^2$, where $v_F=k_F/m^{\star}$ is the Fermi velocity, the orbital critical field is $H_{c20}\simeq\phi_0/2\pi\xi_0^2$ while $H_p=\Delta_0/\sqrt{2}\mu$ is the Pauli-limiting field at zero temperature. 

\section{The critical field temperature derivative} The GL superconducting free energy density has the form
\begin{equation}\label{eq:functional} 
{\cal F}_s
=
{\cal F}_{n0}+
\frac{{\bf h}^{2}}{8\pi}
+\alpha|\Delta|^{2}+\beta|\Delta|^{4} 
+\gamma|{\bf D}\Delta|^{2}.  
\end{equation}
Here, ${\cal F}_{n0}$ is the free energy density
in normal state in absence of magnetic field, ${\bf D}=-i{\bf \nabla}+2e{\bf A}$, 
${\bf h}=\text{rot}{\bf A}$ is the local internal magnetic field, the  induction $B$ determined  by the spatial average 
$\overline{{\bf h}}\equiv{\bf B}=B\hat{z}$ and  the coefficient 
\begin{equation}
\alpha=\alpha_0\frac{T-T_c}{T_c}+ a \left (\frac{\mu B}{T_c}\right )^2
\label{alpha}
\end{equation}
includes the paramagnetic depairing effect.
The solution of the  linearized GL equation as  the linear combination of Landau wave functions with level $n=0$  yields the equation for the upper critical field 
\begin{equation}
\alpha_0\frac{T-T_c}{T_c}+ a \left (\frac{\mu B}{T_c}\right )^2+2e\gamma B=0,
\label{H_c}
\end{equation}
where $T_c$ is the critical temperature at zero field.  This formula is valid for a type of superconducting state with singlet pairing  and one component order parameter in a metal with a form of the Fermi surface. The value of coefficients can, of course, have different  values 
in concrete materials  with different purities for a concrete field orientation in respect to crystallographic axes. For a reader convenience we point out  here 
their values valid near $T_c$ for a clean
 $s$-wave superconconductor with spherical Fermi surface
$\alpha_0=N_0$,         $a=\frac{7\zeta(3)N_0}{4\pi^2}$, $\beta=\frac{7\zeta(3)N_0}{16\pi T_c^2}  $, $\gamma =\frac{7\zeta(3)N_0v_F^2}{32\pi^2T_c^2}$. Here, $N_0$ is the density of states at Fermi level  and we put  $k_B=\hbar=c=1$.  For the more lower temperatures (say at $T\sim T_c/2$) one must use temperature and field dependent  $\alpha, \beta$ and $\gamma$ coefficients \cite{Houzet,Houz06}.

Solving Eq. (\ref{H_c}) we obtain
\begin{equation}
H_{c2}=\frac{e\gamma T_c^2}{a\mu^2}\left [-1+\sqrt{1+\frac{\alpha_0a\mu^2}{(e\gamma T_c)^2}\frac{T_c-T}{T_c}}\right ]
\label{H_{c2}}
\end{equation}
 In the limiting case of pure orbital depairing that is at $\alpha_M<<1$ we obtain from   Eq. (\ref{H_{c2}}) or directly from Eq. (\ref{H_c}) the orbital critical field
\begin{equation}
H_{c2}^{orb}=\frac{\alpha_0}{2e\gamma}\frac{T_c-T}{T_c}.
\label{orb}
\end{equation}
In the opposite case at $\alpha_M>>1$, and  $\frac{T_c-T}{T_c}>1/\alpha_M^2$  the limited by paramagnetic effect 
critical field is 
\begin{equation}
H_{c2}^{p}=\frac{T_c}{\mu}\sqrt{\frac{\alpha_0}{a}\frac{T_c-T}{T_c}}.
\label{p}
\end{equation}
By differentiation of Eq. (\ref{H_c})
we have
\begin{equation}
-\frac{d H_{c2}}{d T}=\frac{\frac{\alpha_0}{2e\gamma T_c}}{1+\frac{a\mu^2}{e\gamma T_c^2}
H_{c2}}
\label{general}
\end{equation}
Substituting here the expression (\ref{H_{c2}}) we obtain
\begin{equation}
-\frac{d H_{c2}}{d T}=\frac{\frac{\alpha_0}{2e\gamma T_c}}{\sqrt{1+ \frac{\alpha_0a\mu^2}{(e\gamma T_c)^2}\frac{T_c-T}{T_c}}}.
\label{final}
\end{equation}
This expression is valid for a type of superconducting state with singlet pairing  and for the superconducting states with triplet but not equal spin pairing states in a metal with a form of the Fermi surface.  For the clean superconductor one can rewrite this as follows
\begin{equation}
-\frac{d H_{c2}}{d T}=\frac{\frac{\alpha_0}{2e\gamma T_c}}{\sqrt{1+ C\alpha^2_M\frac{T_c-T}{T_c}}},
\label{cleanfinal}
\end{equation}
$C$ in the denominator is a constant of the order of unity. 

In the limit of small Maki parameters the critical field temperature derivative is determined only by the orbital effect. It is temperature independent and given by the numerator of Eq. (\ref{final}). While in a superconductor with strong paramagnetic effect that is  at large enough Maki parameters 
 the value of $|dH_{c2}/dT|$ rapidly decreases with decreasing temperature, which leads in its turn to the fast decrease of the Ginzburg-Landau parameter (\ref{GL}).  

\section{Comparison with the paper \cite{Houzet}} We have found the temperature dependence of the Ginzburg-Landau parameter basing on the Ehrenfest relation (\ref{Ehrenfest}).
Meanwhile as we already pointed out there was derived an expression for $\kappa$ valid in the limit of strong  paramagnetic depairing. 
To compare these results it is convenient  begin with the general formula \cite{footnote} for the spacial average of superconducting energy density 
\begin{equation}\label{freered}
\overline{{\cal F}_s}
={\cal F}_{n0}+
\frac{B^2}{8\pi }
-\frac{(\overline{{\cal F}_2(\Delta,{\bf A}_0)})^2}
{4\left [\overline{{\cal F}_4(\Delta,{\bf A}_0)}-\frac{\overline{{\bf h}_1^2}}{8\pi }\right ]}, 
\end{equation}
where ${\cal F}_2$ and ${\cal F}_4$ collect together quadratic and quartic 
terms with respect to $\Delta$, respectively.
Just below the upper critical line defined by $H_{c2}(T)$, the magnetic field is partially
screened by supercurrents and we decompose ${\bf h}={\bf B}+{\bf h}_1$,  
such that  $\overline{{\bf h}_1}=0$, and, correspondingly, 
${\bf A}={\bf A}_0+{\bf A}_1$.

 Starting this formula one can derive general expression for $\kappa$ at arbitrary Maki parameter value. However,  to escape the cumbersome formulae we consider only the situations with  $\alpha_M<<1$ and $\alpha_M>>1$.  In the first case 
\begin{equation}
\overline{{\cal F}_2(\Delta,{\bf A}_0)}=2e\gamma [(B-H_{c2}^{orb}(T)]\overline{|\Delta|^2},
\end{equation}
in the second one 
\begin{equation}
\overline{{\cal F}_2(\Delta,{\bf A}_0)}=\varepsilon[(B-H_{c2}^{p}(T)]\overline{|\Delta|^2}.
\end{equation}
Here,
\begin{equation}
\varepsilon=\left (\frac{\partial \alpha}{\partial B}\right )_{B=H_{c2}^p}
=
\frac{2a\mu^2H_{c2}^p}{T_c^2}
\end{equation}
In any case 
\begin{equation}
\overline{{\cal F}_4(\Delta,{\bf A}_0)}=\beta\beta_A(\overline{|\Delta|^2})^2.
\end{equation}
Then, taking into account the screening currents term $\frac{\overline{\bm{h}_1^2}}{8\pi }$
in denominator of Eq. (\ref{freered}) we come\cite{Houzet} to equation
\begin{equation}
\overline{{\cal F}_s}
={\cal F}_{n0}+
\frac{B^2}{8\pi }
-\frac{(B-H_{c2}(T))^2}
{8\pi[1+\beta_A(2\kappa^2-1)]},
\end{equation}
valid at any the  Maki parameter  value. 
But  at $\alpha_M<<1$ one must put here the upper critical field as determined by Eq. (\ref{orb}) and the Ginzburg - Landau parameter is
\begin{equation}
\kappa=\kappa_{GL}=\frac{\sqrt{\beta}}{4\sqrt{\pi}e\gamma }.
\end{equation}
Whereas at $\alpha_M>>1$ and  $\frac{T_c-T}{T_c}>1/\alpha_M^2$ one must use the upper critical field as determined by Eq. (\ref{p}) and
the Ginzburg - Landau parameter is
\begin{equation}
\kappa=\frac{\sqrt{\beta}}{2\sqrt{\pi}\varepsilon}.
\end{equation}
The latter for a clean superconductor can be rewritten as
\begin{equation}
\kappa\approx\frac{\kappa_{GL}}{\alpha_M\sqrt{\frac{T_c-T}{T_c}}}.
\end{equation}
This expression is in obvious correspondence with Eqs. (\ref{cleanfinal}) and (\ref{GL}).

\section{Conclusion} The derived temperature dependence of the Ginzburg-Landau parameter  is consistent with experimental observations \cite{Haga} in several heavy fermionic superconductors CeCoIn$_5$, URu$_2$Si$_2$, 
NpPd$_5$Al$_2$. In all these compounds the phase transition to the superconducting state becomes of the first order at low temperature - high field region \cite{Izawa, Kasahara, Aoki}, that directly demonstrates the dominant role of  paramagnetic depairing mechanism.

Similar observations have been done recently\cite{Shimizu} in heavy fermionic compound UBe$_{13}$.
This case demands further investigation because it seems that this material having extremely high upper critical field \cite{Glemot} and $ T^3$ behavior
of specific heat at low temperatures \cite{Ott} belongs to triplet superconductors with point nodes
in the quasiparticle spectrum.

\acknowledgements
One of the authors (V. P. Mineev) expresses his  gratitude to Dr. Y. Haga for his interesting talk at Reimei Symposium (Grenoble, February 2012) and for sending many experimental results that has stimulated this simple theoretical reseach.

The work has been supported by the SINUS grant of the Agence Nationale de la Recherche.


\begin{thebibliography}{99}

\bibitem{Abrikosov}A. A. Abrikosov, Sov. Phys. JETP {\bf 5} (1957) 1174.

\bibitem{Ikeda}S. Ikeda, H. Shishido, M. Nakashima, R. Settai, D. Aoki, Y. Haga, H. Harima, Y. Aoki, T. Namiki, H. Sato, and Y. Onuki:
J. Phys. Soc. Jpn. {\bf 70} (2001) 2248.

\bibitem{Haga} Y.Haga: privite communication.

\bibitem{Adachi}H. Adachi, M. Ichioka, and K. Machida: J. Phys. Soc. Jpn. {\bf 74} (2005) 2181.

\bibitem{Houzet} M. Houzet and V. P. Mineev: Phys. Rev. B {\bf 76} (2007)  224508.

\bibitem{Houz06}M. Houzet and V. P. Mineev: Phys.Rev.B {\bf 74}
 (2006) 144522.


\bibitem{footnote} The general expression for the spacially averaged energy is derived exactly in the same manner as Eq. (20) in the paper  \cite{Houzet} where it was done  for  $\alpha_M>>1$.

\bibitem{Izawa} K.Izawa, H. Yamaguchi, Y. Matsuda, H. Shishido, R. Settai, and Y. Onuki: Phys. Rev. Lett. {\bf 87} (2001) 057002.

\bibitem{Kasahara} Y. Kasahara, T. Iwasawa, H. Shishido, T. Shibauchi, K.Behnia, Y. Haga, T. D. Matsuda, Y. Onuki, M. Sigrist, and Y. Matsuda: Phys. Rev. Lett. {\bf 99} (2007) 116402.

\bibitem{Aoki} D. Aoki, Y. Haga, T. D. Matsuda, N. Tateiwa, S. Ikeda, Y. Homma, H. Sakai, Y. Shiokawa, E. Yamamoto, A. Nakamura, R. Settai,  and Y. Onuki: J. Phys. Soc. Jpn. {\bf 76} (2007) 063701.


\bibitem{Shimizu}
Y. Shimizu, Y. Ikeda, T. Wakabayashi, Y. Haga, K. Tenya, H. Hidaka, T. Yanagisawa, and H. Amitsuka: \\J. Phys. Soc. Jpn. {\bf 80} (2011) 093701.
 
 \bibitem{Glemot} L. Glemot, J. P. Brison,  J. Flouquet, A. I. Buzdin, I. Sheikin, D. Jaccard, C. Thessieu, and F. Thomas:  Phys. Rev. Lett. {\bf 82} (1999) 169.
 
 \bibitem{Ott} H. R. Ott, H. Rudiger, Z. Fisk, and J. L. Smith: Phys. Rev. Lett. {\bf 50} (1983) 1595.
\end{thebibliography}
\end{document}